\documentstyle[12pt]{article}

\newcommand{\be}{\begin{equation}}
\newcommand{\ee}{\end{equation}}
\newcommand{\bea}{\begin{eqnarray}}
\newcommand{\eea}{\end{eqnarray}}

\begin{document}
\begin{titlepage}


\vspace{1in}

\begin{center}
\Large
{\bf Separable Brane Cosmologies in Heterotic M--Theory}

\vspace{1in}

\normalsize

\large{James E. Lidsey$^1$}

\normalsize
\vspace{.7in}
 
{\em Astronomy Unit, School of Mathematical 
Sciences,  \\ 
Queen Mary \& Westfield, Mile End Road, LONDON, E1 4NS, U.K.}

\end{center}

\vspace{1in}

\baselineskip=24pt
\begin{abstract}
\noindent It is shown that any anisotropic 
and inhomogeneous cosmological solution to the 
lowest--order, four--dimensional, dilaton--graviton 
string equations of motion may be 
employed as a seed to derive
a curved, three--brane cosmological 
solution to five--dimensional 
heterotic M--theory compactified on a Calabi--Yau three--fold.
This correspondence formally
relates a weakly coupled string cosmology directly with a 
strongly coupled one. 
The asymptotic behaviour of a wide class of spatially homogeneous 
braneworlds is deduced. 
Similar solutions may be derived in toroidally 
compactified massive type IIA supergravity.

\end{abstract}

\vspace{.3in}

\vspace{.7in}
$^1$Electronic mail: jel@maths.qmw.ac.uk
 
\end{titlepage}


The ${\rm E}_8\times {\rm E}_8$ heterotic string theory is 
phenomenologically favoured from a particle physics perspective, 
because it may contain the standard model
of gauge interactions \cite{gsw}. 
The strongly coupled regime of this theory has been interpreted 
by Ho\v{r}ava and Witten as M--theory 
on the orbifold ${\rm R}^{10} \times S^1/{\rm Z}_2$ with an  
infra--red limit corresponding to 
eleven--dimensional supergravity on a manifold with two 
ten--dimensional
boundaries \cite{hw}. 
A set of ${\rm E}_8$ gauge supermultiplets 
on each of the orbifold fixed planes ensures the cancellation of 
anomalies. 
Compactification of this theory on a Calabi--Yau three--fold 
results in a gauged, five--dimensional supergravity theory  
with two four--dimensional boundaries \cite{witten,lukas}. The theory does 
not admit five--dimensional Minkowski space as a consistent 
solution. Instead the `vacuum' corresponds to a 
static BPS state given by a warped product
of four--dimensional flat space and an interval \cite{lukas}. 
Such a solution may be interpreted as 
a pair of parallel three--branes with equal and opposite tension 
that are located at the orbifold planes. Our observed 
four--dimensional spacetime is then interpreted as 
the world--volume of one of these branes. 

This interpretation has significant implications 
for early universe cosmology and it is therefore important 
to derive cosmological models in this context. 
Such a study involves a search for time--dependent
solutions to the field equations. Cosmological 
solutions in heterotic M--theory have been found previously by reinterpreting 
temporal and radial coordinates in certain
$p$--brane backgrounds \cite{benakli}. 
An alternative method was adopted by Lukas {\em et al.}, 
who found spatially 
flat Friedmann--Robertson--Walker (FRW) cosmological brane solutions
by compactifying Ho\v{r}ava--Witten theory on a Calabi--Yau 
space \cite{lukascos}. 
This approach was subsequently generalized to the spatially 
curved, 
FRW models  by Reall \cite{reall}. Domain 
wall solutions moving in a time--dependent bulk 
have also been found \cite{cr} and 
a solution preserving $N=1$ supersymmetry in 
four dimensions was recently presented \cite{susysol}. 

It is reasonable to suppose that spatial anisotropies 
and inhomogeneities in 
our three--dimensional subspace
would have been significant in the environment 
of the very early universe and it is therefore important 
to develop models that take these effects into account. 
In view of this, we derive a wide class of 
spatially anisotropic and inhomogeneous brane 
cosmologies in  Ho\v{r}ava--Witten M--theory.  
We consider the class of five--dimensional 
models with an action of the form 
\begin{equation}
\label{5daction}
S=\frac{1}{2} \int d^4 x \int^{+\pi}_{-\pi} 
dy \sqrt{-g} \left[ R -\frac{1}{2}
\left( \nabla \varphi \right)^2 -\frac{\alpha^2}{3} e^{-a \varphi}
\right] + \sum_{i=1}^2 \int_{{\cal{M}}_4^{(i)}} d^4 x\sqrt{-g_i} 
\mu_i e^{-a \varphi /2} ,
\end{equation}
where $R$ is the Ricci scalar of the bulk spacetime
with metric $g_{AB}$, $g \equiv {\rm det}g_{AB}$ and 
$\{ a, \mu_i \}$ are constants\footnote{In this paper, 
the spacetime metric has signature $(-, +, +, 
\ldots , +)$. Upper case, Latin indices without a circumflex accent 
vary from $A=(0, 1, \ldots, 4)$, upper case Latin  
indices with circumflex accents take values in the range 
$\hat{A}=(0, 1, \ldots , 10)$, lower case Greek indices span 
$\mu =(0, 1, 2, 3 )$ and lower case Latin indices 
denote spatial dimensions.}. The 
orbifold, $S^1/{\rm Z}_2$, is parametrized by the 
coordinate, $y$, and  corresponds to a segment 
of the real line bounded by two fixed 
points on the circle. 
It is defined over the 
coordinate range $y \in [ -\pi , \pi ]$, where the 
endpoints are identified and the  discrete 
symmetry $y \rightarrow -y$ is imposed.
This transformation reverses the 
orientation of the circle and the fixed points are 
specified by the conditions $y= \{ 0, \pi \}$. These 
coordinates determine the location  of 
the four--dimensional hyperplanes, ${\cal{M}}_4^{(i)}$. 
The metrics on these orbifold fixed planes are defined by 
$g_{\mu\nu}^{(1)} \equiv g_{\mu\nu}(y=0)$ 
and $g_{\mu\nu}^{(2)} \equiv g_{\mu\nu}
(y=\pi )$, respectively, and $g^{(i)} 
\equiv {\rm det} g^{(i)}_{\mu\nu}$. For consistency, 
we further require that all degrees of freedom 
are invariant under the ${\rm Z}_2$ transformation. 

When $a=2$ and 
$\mu_1 =-\mu_2 = \sqrt{2} \alpha$, 
action (\ref{5daction}) represents a truncation 
of Ho\v{r}ava--Witten theory compactified on a 
Calabi--Yau three--fold  \cite{lukas}, where the 
eleven--dimensional metric 
is given by 
\begin{equation}
\label{11dmetric}
d\hat{s}^2_{11} =e^{-2\varphi /3} 
g_{AB}dx^A dx^B +e^{\varphi /3} \Omega_{mn}dz^mdz^n
\end{equation}
and the Calabi--Yau space has a metric $\Omega_{mn} = 
\Omega_{mn} (z^p)$ 
\cite{lukas}.
The radius of the Calabi--Yau space is parametrized 
by the scalar field, $\varphi$, that self--interacts via a 
Liouville potential. 
This interaction potential arises because 
a non--zero internal component of the four--form field strength  
must be included if the compactification 
from eleven to five dimensions is to be consistent \cite{lukas}. 
This contribution is parametrized by the 
constant, $\alpha$, that specifies the brane tensions. 

In deriving the field equations from the action (\ref{5daction}), 
we assume that the constants $\{ a, \mu_i \}$ 
are arbitrary. The gravitational and moduli 
field equations then take the form 
\begin{eqnarray}
\label{5deinstein}
G_{AB} =\frac{1}{2} \nabla_A \varphi \nabla_B 
\varphi -g_{AB} \left( \frac{1}{4} \left( \nabla \varphi 
\right)^2 + \frac{\alpha^2}{6} e^{-a\varphi} \right) \nonumber \\
+
g_{\mu A}g_{\nu B} e^{-a\varphi /2} \left[ \mu_1 \delta (y) 
g^{\mu\nu}_{(1)} \sqrt{g_1/g}
+\mu_2 \delta (y-\pi ) g^{\mu\nu}_{(2)} \sqrt{g_2/g} \right]
\end{eqnarray}
\begin{equation}
\label{5dvarphi}
\nabla^2 \varphi =- \frac{a\alpha^2}{3} e^{-a\varphi} +
a e^{-a \varphi /2} \left[ \mu_1 
\delta (y)  \sqrt{g_1/g} + \mu_2  
\delta ( y- \pi ) \sqrt{g_2/g} \right]   ,
\end{equation}
respectively. 

We search for cosmological solutions 
where the 
five--dimensional metric has the general form 
\begin{equation}
\label{bucher}
ds^2_5 =H^{4/(3\Delta)} f_{\mu\nu} dx^{\mu}dx^{\nu} 
+H^{16/(3\Delta )}e^{2 \beta} dy^2
\end{equation}
and the constant 
\begin{equation} 
\label{parameter}
\Delta \equiv \frac{3a^2 -8}{3}
\end{equation}
is assumed to be non--zero.  The warp factor, 
$H=H(y)$, depends only on the orbifold coordinate and the 
four--dimensional 
metric, $f_{\mu\nu} = 
f_{\mu\nu}(x^{\lambda})$, and scalar 
function, $\beta =\beta (x^{\mu})$, 
are independent of this variable. 
The components of the five--dimensional 
Einstein tensor for the metric (\ref{bucher}) are
\begin{eqnarray}
\label{munucomponent}
G_{\mu\nu} =\bar{G}_{\mu\nu} -\bar{\nabla}_{\mu\nu} \beta
-\bar{\nabla}_{\mu} \beta \bar{\nabla}_{\nu}\beta +f_{\mu\nu}
\left[ \bar{\nabla}^2 \beta + \left( \bar{\nabla} 
\beta \right)^2 \right]
\nonumber \\
+ \frac{2}{\Delta} \left[ 
\frac{H''}{H} -\left( 1+\frac{4}{3\Delta} \right) \frac{H'^2}{H^2} 
\right] 
e^{-2\beta} H^{-4/\Delta} f_{\mu\nu} \\
\label{muycomponent}
G_{\mu y} = \frac{2}{\Delta} \frac{H'}{H} \bar{\nabla}_{\mu} \beta   \\
\label{yycomponent}
G_{yy} = -\frac{1}{2} \bar{R} H^{4/\Delta} e^{2\beta} +
\frac{8}{3\Delta^2} \frac{H'^2}{H^2}  ,
\end{eqnarray}
where a bar indicates that quantities are constructed from the 
world--volume metric, $f_{\mu\nu}$, and a prime denotes differentiation 
with respect to the orbifold coordinate, $y$. 

In this work, we assume a separable {\em ansatz} for the 
Calabi--Yau modulus field: $\varphi \equiv \varphi_1(x) +\varphi_2 (y)$.
As shown recently by Brecher and Perry \cite{brecher}, there exists a class of 
three--brane 
(domain wall) solutions to Eqs. (\ref{5deinstein}) and 
(\ref{5dvarphi}) for the Ho\v{r}ava--Witten 
theory when $f_{\mu\nu}$ is 
a Ricci--flat metric, $\beta = 
\varphi_1 =0$, $\varphi_2 = 3\ln H$ and $H=1+(\sqrt{2}\alpha 
|y| ) /3$. The metric, 
$f_{\mu\nu}$, may then be interpreted as the 
world--volume of the branes. When the world-volume 
is Minkowski space, the solution reduces to the BPS state of Ref. 
\cite{lukas}. 
We generalize this class of solution such that 
the Calabi--Yau radius and orbifold dimension 
have a non--trivial dependence on the world--volume 
coordinates.

The $(\mu y)$--component of the Einstein field equations 
(\ref{5deinstein}) then reduces to 
\begin{equation}
\label{ymusep}
\frac{H'}{H} \bar{\nabla}_{\mu} \beta = \frac{\Delta}{4}  \varphi_2' 
\bar{\nabla}_{\mu} \varphi_1
\end{equation}
and Eq. (\ref{ymusep}) is solved by 
\begin{eqnarray}
\label{ymusolve}
\varphi_2 =\frac{2a}{\Delta} \ln H \\
\label{betasol}
 \varphi_1 = \frac{2}{a} \beta
\end{eqnarray}
for $a \ne 0$. By 
substituting Eqs. (\ref{munucomponent}), (\ref{yycomponent}), 
(\ref{ymusolve}) and (\ref{betasol}) 
into the Einstein equations (\ref{5deinstein}), it follows that 
the remaining non--trivial 
components then take the form 
\begin{eqnarray}
\label{munusimplified}
\bar{G}_{\mu\nu} - \left( 1+ \frac{2}{a^2} \right) 
\bar{\nabla}_{\mu} \beta \bar{\nabla}_{\nu} \beta 
- \bar{\nabla}_{\mu\nu} 
\beta +f_{\mu\nu} \left[ \bar{\nabla}^2 \beta +\left( 
1+ \frac{1}{a^2} \right) \left( \bar{\nabla} \beta \right)^2 \right] 
\nonumber \\
= \left[ -\frac{2}{\Delta} \frac{H''}{H}  +\frac{1}{H} 
\left( \mu_1 \delta (y) + \mu_2 \delta (y -\pi ) \right) 
+ \frac{1}{\Delta} \frac{H'^2}{H^2} -\frac{\alpha^2}{6H^2} \right]
e^{-2\beta} H^{-4/\Delta} f_{\mu\nu} \\
\label{yysimplified}
\frac{1}{2} \bar{R} -\frac{1}{a^2} \left( \bar{\nabla} 
\beta \right)^2 
= - \left[ \frac{1}{\Delta} \frac{H'^2}{H^2} -\frac{\alpha^2}{6H^2} 
\right] e^{-2\beta} H^{-4/\Delta} ,
\end{eqnarray}
where we have employed Eq. (\ref{parameter}). Moreover, 
the field equation (\ref{5dvarphi}) 
for the Calabi--Yau radius simplifies to
\begin{eqnarray}
\label{varphisimplified}
\bar{\nabla}^2 \beta + \left( \bar{\nabla} \beta 
\right)^2 = -\frac{a^2}{2} e^{-\beta} H^{-4/\Delta} 
\left[ \frac{2}{\Delta} \frac{H''}{H} -\frac{1}{H} 
\left( \mu_1 \delta (y) + \mu_2 \delta (y -\pi ) \right) \right. 
\nonumber \\
\left. 
-\frac{2}{\Delta} \frac{H'^2}{H^2} +\frac{\alpha^2}{3H^2} \right] 
e^{-2\beta} H^{-4/\Delta}   .
\end{eqnarray}

The left-- and right--hand sides of each of 
Eqs. (\ref{munusimplified})--(\ref{varphisimplified}) can be separated 
by equating them all to zero. This separates terms that depend only on the 
orbifold coordinate from those that are independent of $y$. 
The right--hand sides vanish if  
\begin{eqnarray}
\label{Hsquare}
H'^2 =\frac{\Delta \alpha^2}{6} \\
\label{Hdouble}
H'' = \frac{\Delta}{2} \left( \mu_1 \delta (y)  + \mu_2 \delta (y-\pi ) 
\right) 
\end{eqnarray}
are simultaneously satisfied 
and the left--hand sides are solved by 
\begin{eqnarray}
\label{massless1}
\bar{R}_{\mu\nu} =\bar{\nabla}_{\mu\nu} \beta 
+ \left( 1+ \frac{2}{a^2} \right) \bar{\nabla}_{\mu} \beta 
\bar{\nabla}_{\nu}  \beta \\
\label{massless2}
\bar{\nabla}^2 \beta + \left( \bar{\nabla} \beta \right)^2 =0 .
\end{eqnarray}

The solution to Eq. (\ref{Hsquare}) satisfying the orbifold 
symmetry is \cite{lukas,pope}
\begin{equation}
\label{nonsingle} 
H =1 +m |y|, \qquad m \equiv \sqrt{\frac{\alpha^2 \Delta}{6}}  .
\end{equation}
Differentiating Eq. (\ref{nonsingle}) twice with respect 
to the orbifold coordinate results in two $\delta$--functions\footnote{We
employ the expressions $|y|' =\epsilon (y) -\epsilon 
(y-\pi) -1$ and $|y|'' =2 \delta (y) -2 \delta (y-\pi)$, 
where $\epsilon (y) =1$ if $y \ge 0$ and $\epsilon (y) =-1$ if $y<0$. 
The second 
$\delta$--function arises because $y$ is periodic.}, 
$\delta (y)$ and $\delta (y -\pi )$. Consistency with Eq. 
(\ref{Hdouble}) then implies that the brane tensions must satisfy
$\mu_1 =-\mu_2 = \left( 8\alpha^2/3\Delta \right)^{1/2}$. When 
$a=2$, these are precisely the conditions arising in the compactified 
Ho\v{r}ava--Witten heterotic theory \cite{lukas}. 
The functional form of the $y$--dependent part
of the Calabi--Yau modulus field 
is then directly deduced from Eq. (\ref{ymusolve}). 
Thus, the orbifold--dependent sector of the solution 
represents a pair of parallel three--branes and has the 
functional form of the vacuum solution of Ref. \cite{lukas}. 

The dynamics of the branes is completely determined 
by the solutions to Eqs. (\ref{massless1})--(\ref{massless2}) 
and it only remains to solve this system of equations. When 
the length of the orbifold 
interval is independent of the world--volume coordinates, 
Eq. (\ref{massless1}) implies that $\bar{R}_{\mu\nu} =0$ and we recover 
the class of Ricci--flat branes found by Brecher and Perry \cite{brecher}. 
Further insight may be gained by performing the conformal transformation
\begin{equation}
{^{(s)}}f_{\mu\nu} = \Omega^2 f_{\mu\nu} , \qquad \Omega^2 \equiv 
e^{b \Phi}
\end{equation}
on the world--volume metric, where 
\begin{equation}
\Phi \equiv \frac{\beta}{b-1} , \qquad b \equiv 
1+\frac{a}{\sqrt{4+3a^2}}  .
\end{equation}
Substituting in the separable 
condition (\ref{betasol}) implies that 
Eqs. (\ref{massless1})--(\ref{massless2}) then reduce to 
\begin{eqnarray}
\label{string1}
{^{(s)}}R_{\mu\nu} =- {^{(s)}}\nabla_{\mu\nu} \Phi \\
\label{string2}
{^{(s)}}\nabla^2 \Phi = \left( {^{(s)}}\nabla \Phi \right)^2
\end{eqnarray}

Eqs. (\ref{string1}) and (\ref{string2}) 
represent the lowest--order 
$\beta$--function equations for the 
massless graviton $({^{(s)}}f_{\mu\nu})$ 
and dilaton $(\Phi)$ excitations 
of the heterotic 
string compactified on a static Calabi--Yau manifold 
in the absence of the Neveu--Schwarz/Neveu--Schwarz 
two--form potential. They are valid in 
the perturbative, weakly coupled regime of the theory. 
Given a solution to these equations, we may 
immediately deduce the 
functional form of the five--dimensional brane metric (\ref{bucher})
and the solution may be reinterpreted in an eleven--dimensional 
context by 
substituting Eq. (\ref{bucher}) into Eq. (\ref{11dmetric}). 
We find for the Ho\v{r}ava--Witten case that  
\begin{equation}
\label{elevenstring}
d\hat{s}^2_{11} =H^{-1} e^{-11\Phi /6}
\left[  {^{(s)}}f_{\mu\nu} dx^{\mu}dx^{\nu} \right] 
+H^2 e^{2\Phi /3} dy^2 +He^{\Phi /6} 
\Omega_{mn} dz^mdz^n   ,
\end{equation}
where the warp factor is given by Eq. (\ref{nonsingle}). 

Thus, it follows that {\em any solution 
to the lowest--order, dilaton--graviton 
$\beta$--function equations may be related, after 
appropriate field redefinitions, to a  
curved, three--brane background of heterotic 
M--theory compactified on a Calabi--Yau three--fold}.  
The solution (\ref{elevenstring}) provides a direct relationship between 
a given heterotic string cosmology that is valid in the 
weakly coupled regime of the theory and 
a brane cosmology that is valid in the 
strongly coupled limit. 
The physical separation between the two branes 
becomes larger as the string coupling increases
and varies as $r \propto g_s^{2/3}$, where $g_s^2 \equiv e^{\Phi}$
parametrizes the string coupling in the 
weakly coupled solution. 
The separable condition (\ref{betasol}) 
implies that the radius of the Calabi--Yau space must 
follow the behaviour of this coupling in a well defined way.
For the specific case of the spatially flat, FRW model, the 
weakly coupled solution to Eqs. (\ref{string1}) and (\ref{string2}) 
is given by the `dilaton--vacuum' solution, $a_s \propto 
\eta^{(1\pm \sqrt{3})/2}$ and $e^{\Phi} \propto \eta^{\pm \sqrt{3}}$, 
where $a_s$ represents the string--frame scale factor and $\eta$ is 
conformal time. This solution
forms the basis of the pre-big bang inflationary scenario \cite{pbb} and
we recover the brane cosmology of Ref. \cite{lukascos} 
when this solution is transformed into Eq. (\ref{elevenstring}). 

A further conformal transformation given by
\begin{equation}
\label{einsteinconformal}
\tilde{f}_{\mu\nu} = \Theta^2  {^{(s)}}f_{\mu\nu} , \qquad 
\Theta^2 \equiv e^{-\Phi}
\end{equation}
implies that Eqs. (\ref{string1}) and (\ref{string2}) 
are dynamically equivalent to 
\begin{eqnarray}
\label{massein1}
\tilde{{R}}_{\mu\nu} =\frac{1}{2} \tilde{{\nabla}}_{\mu} \Phi
\tilde{{\nabla}}_{\nu} \Phi \\
\label{massein2}
\tilde{{\nabla}}^2 \Phi =0  .
\end{eqnarray}
Eqs. (\ref{massein1})--(\ref{massein2}) represent the field equations 
for a massless scalar field, $\Phi$, 
that is minimally coupled to Einstein gravity. 
This correspondence is important because the four--dimensional,  
Einstein--massless scalar field 
system has been extensively studied in the literature,
both within the anisotropic and inhomogeneous cosmological 
settings \cite{exact}. This implies that known techniques 
for generating solutions and analyzing the asymptotic 
behaviour and singular nature of cosmological models in standard 
general relativity may be applied 
directly to the class of brane cosmologies derived above. 
 
For example, in the case of the 
spatially homogeneous Bianchi universes, the massless scalar field 
may be interpreted as a stiff perfect fluid, where the 
speed of sound in the fluid is equal to the speed of 
light. A number of exact solutions are known for this effective equation 
of state \cite{exact}. Moreover, 
the generic asymptotic behaviour of many 
of these cosmologies has been established.
In particular, apart 
from a set of measure zero, all 
orthogonal Bianchi class B cosmologies 
(types IV, V, ${\rm VI}_h$ and ${\rm VII}_h$) containing 
a stiff perfect fluid, where the fluid velocity is orthogonal to 
the surfaces of homogeneity, 
are asymptotic in the future to a plane wave state
and asymptotic in the past to the Jacobs Bianchi type I solution 
\cite{hewitt}. 

The four--dimensional 
Jacobs type I metric \cite{jacobs} is the generalization of the vacuum Kasner 
\cite{kasner} solution 
to include a stiff perfect fluid. In a string cosmological 
context, it is conformally equivalent to the `rolling radii' solution 
found by Meuller \cite{mueller}. The string--frame metric
has the form 
${^{(s)}}ds_4^2 =-dt_s^2 + \sum_{i=1}^3 
t_s^{2p_i} dx_i^2$ 
and the dilaton field is given by $e^{-\Phi} \propto t_s^p$, 
where 
$\sum_{i=1}^3 p_i^2 =1$ and $\sum_{i=1}^3 p_i = 1-p$. 
In general, there is a curvature singularity in the four--dimensional 
metric at $t_s=0$. Since the 
five--dimensional Ricci scalar is given by 
$R= H^{-1} \bar{R} + \ldots$, the solution also admits a five--dimensional 
singularity. Indeed, a similar conclusion holds in eleven
dimensions because, in general, the square of the Riemann tensor 
varies as $\hat{R}_{\hat{A}\hat{B}\hat{C}\hat{D}}
\hat{R}^{\hat{A}\hat{B}\hat{C}\hat{D}} \propto t^{-4}_s$. 
Indeed, in the absence of any dependence on the orbifold coordinate, 
the line--element (\ref{elevenstring}) in the 
early--time limit of these models would correspond 
to a specific Kasner solution of eleven--dimensional, vacuum 
Einstein gravity. 

In principle, strongly coupled, spatially inhomogeneous 
brane cosmologies may also be derived in the manner outlined above. 
The simplest class of inhomogeneous cosmologies that 
generalizes the Bianchi universes are the diagonal 
Einstein--Rosen $G_2$ models \cite{einsteinrosen,kr}. 
These admit two commuting 
spacelike Killing vectors and have a metric of the 
general form $ds^2 = e^{2f(t,x)}(-dt^2+dx^2) + 
t(e^{p(t,x)} dy^2 +e^{-p(t,x)}dz^2 )$, where $\{ f, p \}$ are
scalar functions. Thus, spatial homogeneity is broken along 
the $x$--direction. Numerous techniques exist
for generating a minimally 
coupled, scalar field $G_2$ solution from a corresponding 
vacuum solution of this form. (For a review, see, e.g., 
Ref. \cite{kr}). This class of models is interesting 
because there exists a long standing conjecture that they 
represent the leading--order approximation to the general solution
of Einstein gravity in the vicinity of the 
curvature singularity \cite{bk}. 
An investigation into 
string cosmologies of this type is therefore important and 
a number of recent studies appropriate to 
the weakly coupled regime have recently been presented  \cite{weakG2}. 
It would be interesting to perform a detailed investigation 
into the singular nature of the corresponding strongly coupled models, 
although this 
is beyond the scope of the present work. 
It would also be interesting to further generalize the analysis 
to include cosmologies with only a single isometry. 
These models could be generated from vacuum $G_2$ backgrounds 
by employing the algorithm recently developed by Lazkoz 
\cite{lazkoz}. 

Brane cosmologies may also be found in 
other supergravity theories in the manner outlined in this 
work. A particular example is Romans'
massive 
type IIA theory in ten dimensions \cite{romans}.
This theory represents the low--energy limit of the type IIA superstring
\cite{pol}. 
After a suitable truncation, 
the toroidal compactification of this theory to five 
dimensions results in an effective bulk action 
given by the five--dimensional part of Eq.  (\ref{5daction}), 
where $a^2=20/3$ $(\Delta =4)$ \cite{cowdall}. Indeed, this value 
for the coupling parameter arises 
in all massive supergravity theories derived from the 
Scherk--Schwarz compactification 
of a higher--dimensional theory containing an axion field
\cite{cowdall}\footnote{In the Scherk--Schwarz 
compactification \cite{scherk}, the axion is assumed to have a linear 
dependence on an internal coordinate. This is manifested in the 
lower--dimensional theory as an 
exponential self--interaction potential for 
the modulus field. The reader is referred 
to Ref. \cite{cowdall} for details.}.

In conclusion, therefore, we have shown that Ho\v{r}ava--Witten 
theory admits a wide class of anisotropic 
and inhomogeneous cosmological solutions if the Calabi--Yau radius 
and orbifold dimension scale in a specific way. We 
have directly related these
strongly coupled backgrounds to 
weakly coupled, four--dimensional string cosmologies.  
These models therefore provide a dynamical mechanism 
for interpolating between the weakly  and strongly coupled limits 
of the ${\rm E}_8 \times {\rm E}_8$ heterotic string theory. 

\vspace{.3in}

The author is supported by the Royal Society. We thank
D. Clancy, R. Lazkoz and G. Pollifrone for helpful discussions.

\end{document}